\documentclass[aps,amsmath,amssymb,prl,twocolumn,showpacs]{revtex4}
\usepackage{bm}
\usepackage{epsfig}

\begin{document}

\preprint{draft}

\title{One-dimensional Disordered Density Waves and
  Superfluids:\\
The Role of Quantum Phase Slips and Thermal Fluctuations }

\author{Andreas Glatz}
\author{Thomas Nattermann}
\affiliation{Institut f\"ur Theoretische Physik, Universit\"at zu K\"oln\\
Z\"ulpicher Str. 77, D-50937 K\"oln, Germany}

\date{\today}

\begin{abstract}
  The low temperature phase diagram of $1$D disordered quantum systems
  like charge or spin density waves, superfluids and related systems
  is considered by a full finite T renormalization group approach,
  presented here for the first time.  At zero temperature the
  consideration of quantum phase slips leads to a new scenario for the
  unpinning (delocalization) transition. At finite T a rich
  cross-over  diagram is found which reflects the zero
  temperature quantum critical behavior.

\end{abstract}


\maketitle

The collective behavior of condensed modulated structures like charge
or spin density waves (CDWs/SDWs) \cite{Gruener,Brazovski99}, flux
line lattices (FLLs) \cite{Blatter,NatSchei} and Wigner crystals
\cite{Brazovski99} in random environments has been the subject of
detailed investigations since the early 70th. These are motivated by
the drastic influence of disorder: without pinning CDWs would be ideal
superconductors whereas type-II superconductors would show a finite
resistivity.  In three dimensional systems the low temperature phase
of these structures is determined by a zero temperature disorder fixed
point resulting in quasi-long range order and a glassy dynamics
\cite{Blatter,NatSchei}.  In two dimensions this fixed point is
extended to a fixed line which terminates at the glass transition
temperature \cite{CardyOst,ViFer84}.  In the low temperature phase
correlations decay slightly faster than a power law and the linear
resistivity vanishes (for a recent review see \cite{NatSchei}). In one
dimension the situation is different: the glass temperature is shifted
to $T=0$.  Nevertheless there remains a residual trace of the disorder
which is reflected in the low temperature behavior of spatial
correlations and the dynamics \cite{Feigel80}. 

Clearly, at low temperatures also quantum fluctuations have to be
taken into account. Disorder and quantum fluctuations in $1$D CDWs at
zero temperature have been considered previously (see e.g.
\cite{Fukuyama84,GiaSchulz87}) and an unpinning transition as a
function of the interaction strength was found. Finite temperature
effects were partially incorporated by truncating the renormalization
group (RG) flow at the de Broglie wave length of the phason
excitations\cite{GiaSchulz87}. However, for a complete study of the
thermal to quantum cross-over quantum and thermal fluctuations have to
be considered on an equal footing \cite{Chakra}, which will be the
first goal of this paper.  The second goal will be the consideration
of quantum phase slips which trigger their own quantum phase
transition to the disordered phase and influence also the unpinning
transition.  
Experimentally, 1D behavior can be seen in real
materials e.g. in whiskers with a transverse extension smaller than
the correlation length or in chain like crystals with weak interchain
coupling. In the latter case there is a large cross-over length scale
up to which 1D behavior can be observed \cite{Gruener,Brazovski99}.
The results obtained for the CDWs/SDWs have a 
number of further applications on disordered quantum systems: they
relate e.g. to  the localization transition of Luttinger liquids
\cite{Fukuyama84,GiaSchulz87}, tunnel junction chains \cite{Korshunov}, superfluids \cite{FisherGrinstein},
Josephson coupled chains of these systems if the coupling is treated
in mean-field theory \cite{Fukuyama84}.
In most parts of this paper, we will use the terminology of
CDWs.

Well below the mean field condensation temperature T$_{\text{MF}}$  of the
CDW, the electron density $\rho(x)$ can be written in the
form~\cite{Gruener}
\begin{equation}\label{eq.cdwdensity}
\rho(x)=\rho_0(1+Q^{-1}\partial_x\varphi)+\rho_1\cos[p(\varphi+Qx)]+\ldots
\end{equation}
where $Q=2k_F$, $k_F$ is the Fermi--momentum, $\rho_0$ the mean
electron density and $\rho_1$ is proportional to the amplitude of the
complex (mean field) order parameter $\Delta e^{i\varphi}\sim \langle
b_{Q}+b^+_{-Q}\rangle$.  $b^+_k, b_k,$ denotes the phonon creation
and annihilation operator, respectively.  $\varphi(x)$ is a slowly
varying phase variable. Neglecting fluctuations in $\Delta$, the
Hamiltonian of the CDW is given by
\begin{eqnarray}
 \hat {\cal H} & = &
\int\limits_{0}^{L}\Bigg\{\frac{c}{2}\big[\left(\frac{v}{c}
\right)^2\hat P^2+(\partial_x\hat\varphi)^2\big] +
\sum_i U_{i}\rho(x)\delta(x-x_i)\nonumber\\
& &+W\cos  \Big(\frac{q\pi}{\hbar}\int^x dy\hat P(y)\Big)
\Bigg\}dx\label{eq.qham}
\end{eqnarray}
where $[\hat
P(x),\hat\varphi(x^{\prime})]=\frac{\hbar}{i}\delta(x-x^{\prime})$.
$c=f\frac{\hbar v_F}{2\pi}$ denotes the elastic constant, $v_F$ the
Fermi--velocity, $v$ the effective velocity of the phason excitations
and $f(T)$ the condensate density \cite{Gruener}.  Note that $f(T)$
and $\Delta(T)$ vanish at T$_{\text{MF}}$ whereas $v$ remains finite.
The third term results from the effect of impurities of random
potential strenght $U_i=\pm U_{imp}$ and position $x_i$ and includes a
forward and a backward scattering term proportional to $\rho_0$ and
$\rho_1$, respectively. We will assume, that the mean impurity
distance $l_{imp}$ is large compared with the wave length of the CDW
and that the disorder is weak, i.e. $1\ll l_{imp}Q \ll
c/(U_{imp}\rho_1) $. In this case the Fukuyama--Lee length
${L_{\text{FL}}}=(c/(Up^2))^{2/3}$ is large compared with the impurity
distance, here $U=U_{imp}\rho_1/{\sqrt {l_{imp}}}$. The fourth term in
(\ref{eq.qham}) describes the influence of quantum phase slips by
$\varphi=\pm q\pi$ and will be further discussed below
\cite{FisherGrinstein}. The model (\ref{eq.qham}) includes the four
dimensionless parameters $t=T/\pi\Lambda c$, $K=\hbar v /\pi c$,
$u^2=U^2/\Lambda^3\pi c^2$ and $w=W/\pi c\Lambda^2$, which measure the
strength of the thermal, quantum, and disorder fluctuations and the
probability of phase slips, respectively.  $\Lambda=\pi/a$ is a
momentum cut-off. Although for CDWs and SDWs $K$-values of the order
$10^{-1}$ and $1$, respectively, have been discussed at $T=0$
\cite{Maki}, the expressions relating $K$ and $t$ to the microscopic
theory lead to the conclusion that both diverge by approaching
$T_{\text{MF}}$ whereas the ratio $K/t$ remains finite.  The classical
region of the model is given by $K\ll t$ which can be rewritten as the
condition, that the thermal de Broglie wave length
$\lambda_T=K/(t\Lambda)$ of the phason excitations is small compared
with the lattice spacing $a$.

In order to determine the phase diagram we adopt a standard
Wilson-type renormalization group calculation, which starts from a
path integral formulation of the partition function corresponding to
Hamiltonian (\ref{eq.qham}) with $u,w\ll 1$. We begin with the renormalization of the
disorder term and put $w=0$ for the moment.  The system is transformed
into a translationally invariant problem using the replica-trick.
Going over to dimensionless spatial and imaginary time variables,
$\Lambda x \rightarrow x$ and $\Lambda v\tau \rightarrow \tau$, the
replicated action is given by ($\sigma=(u\rho_0\Lambda/\rho_1Q)^2$)
\begin{eqnarray}
\frac{S^{(n)}}{\hbar} & = & \frac{1}{2\pi K}\sum_{\alpha,\beta}
\int\limits_0^{L\Lambda}dx\int\limits_0^{K/t}d\tau\Bigg\{
\Big[(\partial_x\varphi_{\alpha})^2+(\partial_{\tau}\varphi_{\alpha})^2\Big]
\delta_{\alpha\beta}\nonumber\\
&-& \frac{1}{2K}\int\limits_0^{K/t}d\tau^{\prime}\Big[u^2\cos{p}
\Big(\varphi_{\alpha}(x,\tau)-\varphi_{\beta}(x,\tau^{\prime})\Big)\nonumber\\
&+& \sigma\partial_x\varphi_{\alpha}(x,\tau)\partial_x\varphi_{\beta}(x,\tau^{\prime})\Big]\Bigg\}.
\label{eq:S^n/hbar}
\end{eqnarray}
Integrating over the high momentum modes of $\varphi(x,\tau)$ in a
momentum shell of infinitesimal width $2\pi/b\le|q|\le 2\pi $ but
arbitrary frequencies and  rescaling $x\rightarrow
x'=x/b$, $\tau \rightarrow \tau'=\tau/b$ we obtain the following
renormalization group flow equations (up to one loop):
\begin{eqnarray}
\frac{dK}{dl} & = & -  \frac{1}{2}p^4u^2KB_0(p^2K,\frac{K}{2t})
\coth \frac{K}{2t}\label{eq:dK/dl},\\
\frac{du^2}{dl} & = & \Big[3  -  \frac{p^2K}{2}\coth{\frac{K}{2t}}\Big]u^2, \qquad \frac{dt}{dl}  = t, \label{eq:du/dl}
\end{eqnarray}
\begin{equation}
B_i(\nu,y) = \int\limits_0^yd\tau\int\limits_0^\infty\frac{ dx
  g_i(\tau,x)\cosh{(y-\tau)}(\cosh{y})^{-1}}{\left[1+\left(\frac{y}{\pi}\right)^2(\cosh{\frac{\pi x}{y}}-\cos{\frac{\pi\tau}{y}})\right]^{\nu/4}},\label{B}
\end{equation}
where $l=\ln{b}$ and $ g_0(\tau,x)=\delta (x)\tau^2$.
Note that $B_0(p^2K,\frac{K}{2t})\rightarrow 0$ for $K \rightarrow 0$.

The equation for the flow of $\sigma$ is more involved and will not be
discussed here since it does not feed back into the other flow
equations.  Indeed, we can get rid of the forward scattering term by
rewriting $\hat\varphi(x)=\hat\varphi_b(x)+ \varphi_{f}(x)$ with
$\varphi_{f}(x)=\int^{x}_0 dy c(y)$, $\langle c(x)\rangle=0$ and
$\langle c(x)c(x')\rangle =\frac{\pi}{2}\sigma \delta (x-x')$. The
phase correlation function $C(x,\tau)=\langle
(\varphi(x,\tau)-\varphi(0,0))^2\rangle=C(x,\tau)+C_f(x)$ has
therefore always a contribution $C_f(x)\sim |x|/\xi_f$ with
$\xi_f^{-1} \sim \sigma(l=\log|x|)$.  Since all further remarks about
phase correlations refer to $ C_b(x,\tau)$ we will drop the subscript $b$.

There is no renormalization of $t$ (i.e. $c$) because of a statistical
tilt symmetry \cite{Schultz88}.  The special case $t=0$ was previously
considered in \cite{GiaSchulz87} (with $p=\sqrt 2$).  The flow
equation for $K$ obtained in \cite{GiaSchulz87} for $w=t=0$ deviates
slightly from (\ref{eq:dK/dl}), which can be traced back to the
different RG-procedures. The critical behavior is however the same:
there is a Kosterlitz-Thouless (KT) transition \cite{Kosterlitz} at
$K_u$ between a disorder dominated pinned and a free unpinned phase
which terminates in the fixed point $K_u^*=6/p^2$. $u_0$ denotes the
bare value of the disorder and $K_u$ is given by
$u^2=\frac{K_u^*}{p^2\eta}(\frac{K_u-K_u^*}{K_u^*}-\log{\frac{K_u}{K_u^*}})$
with $\eta=B_0(p^2K_u^*,\infty)$. In the pinned phase the parameters
$K,u$ flow into the classical, strong disorder region: $K\rightarrow
0, u\rightarrow \infty$.  The correlation function $C(x,0)\sim
|x|/\xi_u$ increases linearly with $|x|$.  Integration of the flow
equations gives for small initial disorder and $K\ll K_u$ an
effective correlation length $\xi_{u} \approx
\Lambda^{-1}(\Lambda L_{\text{FL}})^{(1-K/K_u)^{-1}} $ at which $u$
becomes of the order unity.  Close to the transition line $\xi_u$
shows KT behavior.  For $K \ge K_u$ $\xi_u$ diverges and
$C(x,\tau)\sim K(l=\log|z|)\log|z|$ where $|z|=\sqrt{x^2+\tau^2}$.
Note that $K(l)$ saturates on large scales at a value
$K_{\text{eff}}(u_0)$.

For large values of $u$ our flow equations break down, but we can find
the asymptotic behavior in this phase by solving the initial model in
theit {\it strong pinning limit} $U_{\text{imp}}\gg 1, K=0$  exactly.
A straightforward but somewhat clumsy calculation yields for the pair
correlation function $C(x,\tau)=\frac{2\pi}{p\alpha}(1-
\frac{\alpha}{\sinh\alpha})|Qx|$ where $\alpha=\pi /(pQl_{\text{imp}})$.
 The connection to the weak pinning model follows by
choosing $l_{\text{imp}}\approx L_{\text{FL}} $.

At {\it finite temperatures} thermal fluctuations destroy the quantum
interference effects which lead to the pinning of the CDW at $t=0$.
The RG flow of $u$ in the region $K < K_u$ first increases and then
decreases. $\xi $ can be found approximately by integrating the flow
equations until the maximum of $u(l)$ and $t(l)/(1+K(l))$ is of 
order one.  This can be done in full generality only numerically (see
Fig. 1). It is however possible to discuss several special cases
analytically.  The zero temperature correlation length can still be
observed as long as this is smaller than $\lambda_T$ which rewrites
for $K$ not too close to $K_u^*$ as $t\lesssim t_K\approx
Kt_{\text{u}}^{(1-K/K_u)^{-1}}$, $t_{u}\approx (\Lambda
L_{\text{FL}})^{-1}$. We call this domain the {\it quantum disordered
  region}.  For $K\ge K_u$ the correlation length $\xi$ is given by
$\lambda_T$ which is larger than given by purely thermal fluctuations.
For scales smaller than $\lambda_T$, $C(x,\tau)$ still increases as
$\sim \log |z|$ with a continuously varying coefficient
$K_{\text{eff}}(u_0)$. In this sense one observes {\it quantum critical
  behavior} in that region, despite of the fact, that the correlation
length is now finite for all values of $K$ \cite{Chakra}.  In the {\it
  classical disordered region} $t_K<t<t_{u}$ the correlation length
is roughly given by $L_{\text{FL}}$ as follows from previous studies
\cite{Feigel80,ViFer84}. In the remaining region $t_{u}\lesssim t$
we adopt an alternative method by mapping the (classical)
one-dimensional problem onto the Burgers equation with noise
\cite{HuseHenleyFi85}. In this case the RG-procedure applied to this
equation becomes trivial since there is only a contribution from a
single momentum shell and one finds for the correlation length
$\xi^{-1}\approx \frac{\pi}{2}f(T)t(1+(2\pi/p)^2(t_{u}
/t)^3)\Lambda $.  The phase diagram depicted in Fig. 1 is the result
of the numerical integration of our flow equations and shows indeed
the various cross-overs discussed before.

So far the phase field was considered to be single valued. Taking
into account also amplitude fluctuations of the order parameter
the phase may change by multiples of $2\pi$ by orbiting (in space
and imaginary time) a zero of the amplitude. Such vortices
correspond to {\it quantum phase slips} described by the last term in
(\ref{eq.qham}) (with $q=2$), which we discuss here under
equilibrium conditions.  This operator superposes two translations
of $\varphi$ by $\pm q\pi$ left from $x$, i.e.  it changes
coherently the phase by $\pm q\pi$ in a macroscopic region.  For
vanishing disorder the model can be mapped on the sine-Gordon
Hamiltonian for the $\theta$-field (with $K$ replaced by $K^{-1}$)
by using the canonical transformation $\hat P = -
\frac{\hbar}{\pi}\partial_x\hat \theta$ and $ -
\frac{\hbar}{\pi}\partial_x\hat \varphi=\hat \Pi $.  To see the
connection to space-time vortices one rewrites the action of
interacting vortices as a classical $2D$ Coulomb gas which is
subsequently mapped to the sine--Gordon model \cite{Jose.et.al}.
The initial value $w_0$ of $w$ is proportional to the fugacity
$w_0\approx e^{-S_{\text{core}}/\hbar}$ of the space-time vortices
which may be non-negligible close to T$_{\text{MF}}$ , where the
action $S_{\text{core}}\approx \hbar/(\pi K)$ of the vortex core
is small. Performing an analogous calculation as before (but with
$u=0$) the RG-flow equations read:
 \begin{eqnarray}
    \frac{dK}{dl} & = & -\frac{\pi}{2}\frac{q^4w^2}{K^3}
   B_2\Big(\frac{q^2}{K},\frac{K}{2t}\Big)\coth{\frac{K}{2t}},
\label{eq:dK/dl.sg}\\
\frac{dt}{dl} & = & \left[1-\frac{\pi}{2}\frac{q^4w^2}{K^4}
   B_1\Big(\frac{q^2}{K},\frac{K}{2t}\Big)\coth{\frac{K}{2t}}\right]t,
   \label{eq:dt/dl.sg}\\
   \frac{dw}{dl} & = & \left[2-\frac{q^2}{4K}\coth{\frac{K}{2t}}\right]w,
   \label{eq:dw/dl.sg}
     \end{eqnarray}
     where $B_{1,2}$ are given in (\ref{B}) with $g_1=2\tau^2\cos{x}$
     and $g_2=(x^2+\tau^2)\cos{x}$.  From (\ref{eq:dK/dl.sg}) -
     (\ref{eq:dw/dl.sg}) we find, that for $t=u=0$ quantum phase slips
     become relevant (i.e. $w$ grows) for $K >K_w$ with
     $K_w^*=q^2/8$ ($q=2$ for CDWs). In this region vortices destroy
     the quasi long range order of the CDW, $C(x,\tau)\sim
     |z|/\xi_{w}$.  The transition is of KT type with a correlation
     length $\xi_{w}$ ($w(\log \xi_{w})\approx 1$) diverging at $K_w+0$
     \cite{FisherGrinstein}.  At finite temperatures $w$ first
     increases, but then decreases and flows into the region of large
     $t$ and small $w$. Thus quantum phase slips become irrelevant at
     finite temperatures.  This can be understood as follows: at
     finite $t$ the 1D quantum sine-Gordon model can be mapped on the
     Coulomb gas on a torus of perimeter $K/t$ since periodic boundary
     conditions apply now in the $\tau$-direction.  Whereas the
     entropy of two opposite charges increases for separation $L\gg
     K/t$ as $\log (LK/t)$, their action increases linearly with $L$.
     Thus, the charges remain bound.  The one-dimensional Coulomb gas
     has indeed only an insulating phase \cite{Lenard}.
\begin{figure}[h]
\includegraphics[width=0.8\linewidth]{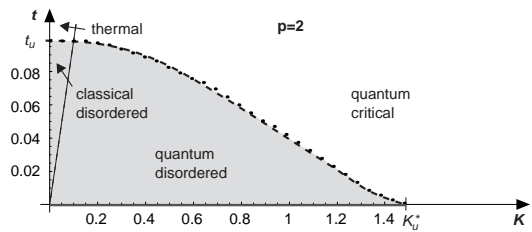}
\caption{The low temperature cross-over diagram of a
one-dimensional
  CDW. The amount of disorder corresponds to a reduced temperature
  $t_{\text{u}}\approx 0.1$. In the classical and quantum disordered
    region, respectively, essentially the $t=0$ behavior is seen. The
    straight line separating them corresponds to $\lambda_T\approx a$.
    In the quantum critical region the correlation length is given by
$\lambda_T$. Pinning (localization) occurs only for $t=0,
K<K_u^*$.}\end{figure}
\begin{figure}[h]
\includegraphics[width=0.8\linewidth]{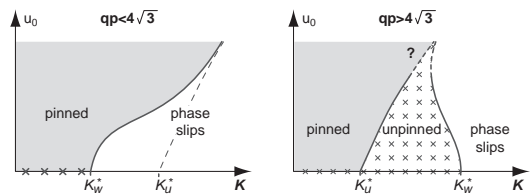}
\caption{$T=0$ phase diagram for a CDW with quantum phase slips.
If
 $qp<4\sqrt{3}$ there is a single transition between a low-K pinned
and a high-K unpinned phase. In both phases the correlation length
is finite. If   $qp>4\sqrt{3}$ these two phases are separated by
a third phase in which phase slips are suppressed and
$C(x,\tau)\sim \log |z|$. Both transitions disappear at finite
$t$.}
\end{figure}

     It is now interesting to consider the combined influence of
     disorder and phase slips.  In doing this we write an approximate
     expression for the action of a single vortex in a region of
     linear extension $L$ as
\begin{equation}
\frac{S_{\text{vortex}}- S_{\text{core}}}{\hbar}= (\frac{q^2}{4K}-2)\log L
- \frac{u(L)}{K}.
\label{Svortex}
\end{equation}
 For very low $K (<K_u, K_w)$ where
$u(L)\approx u_0L^{3/2}$ the disorder always favours vortices on
the scale of the effective Fukuyama-Lee length $\xi_{\text{u}}$.
These vortices will be pinned in space by  disorder. On the other
hand, for very large values of $K (>K_u, K_w)$ phase vortices are not
influenced by disorder since $u(L)$ is renormalized
to zero.  In the remaining region we have to distinguish the cases
$K_u\gtrless K_w$.  For $K_w<K<K_u$ (i.e.  $qp<4\sqrt{3}$) and $u_0=0$
the phase correlations are lost on the scale of the KT correlation
length $\xi_w$ of the vortex unbinding transition. Not too close to
this transition $\xi_w\Lambda\approx
e^{(S_{\text{core}}/2\hbar)(1-{K_v}/{K})^{-1}}$.  Switching on the
disorder, u will be renormalized by  strong phase fluctuations which
lead to an exponential decay of $u\sim u_0
e^{-{\text{const}}L/\xi_w}$ such that disorder is irrelevant for the
vortex gas as long as $\xi_w \lesssim \xi_{\text{u}}$. We expect
that the relation $\xi_w\approx \xi_{\text{u}}$ determines the
position of the phase boundary between a pinned low-$K$ phase where
vortices are favoured by the disorder and an unpinned high-$K$ phase
where vortices are induced by quantum fluctuations. This line
terminates in $K_w$ for $u_0 \rightarrow 0$ (see Fig. 2). If
$S_{\text{core}}$ is large, $\xi_w $ will be large as well and 
$\xi_w\approx \xi_{\text{u}}$ will be reached only for $K\approx
K_u^*$.  For moderate values of $S_{\text{core}}$, the unpinning
transition may be lowered considerably by quantum phase slips.  In the
opposite case $K_u<K<K_w$ (i.e. $qp>4\sqrt{3}$) phase fluctuations
renormalize weak disorder to zero such that vortices are still
suppressed until $K$ reaches $K_w$ where vortex unbinding occurs.  In
this case two sharp phase transitions have to be expected.

Our flow equations describe also the
effect of a {\it commensurate lattice potential} on the CDW: if the
wave length $\pi/k_F$ of the CDW modulation is commensurate with the
period $a$ of the underlying lattice such that $\pi/k_F=n/(qa)$ with
$n,q$ integer, an Umklapp term $w\cos q\varphi$ appears in the
Hamiltonian \cite{Gruener}. We obtain the results in this case from
(\ref{eq:dK/dl.sg}) - (\ref{eq:dw/dl.sg}) (and the conclusions derived
from them) if we use the replacements $K\rightarrow K^{-1},
t\rightarrow t/K^2$ and $w\rightarrow w/K^2$. Thus the lattice
potential is relevant for $K<K_w$ with $K_w^*=8/q^2$.

Next we consider the application of the results obtained so far to a
{\it one-dimensional Bose fluid}.  Its density operator is given by eq.
(1) if we identify $Q/\pi=\rho_0=\rho_1$ ($p=2$).  $\partial_x\varphi
$ is conjugate to the phase $\theta$ of the Bose field \cite{Haldane}.
With the replacements $K\rightarrow K^{-1}$, $t\rightarrow t/K^2$ and
$w=0$, (\ref{eq:S^n/hbar}) describes the action of the 1D-superfluid
in a random potential. $v$ denotes the phase velocity of the sound
waves with $v/(\pi K)=\rho_0/m$ and $\pi v K=\kappa/(\pi^2\rho_0^2)$
where $\kappa$ is the compressibility. The transition between the
superfluid and the localized phase occurs for $K_u^*=2/3$
\cite{GiaSchulz87}.  Thermal fluctuations again suppress the disorder
and destroy the  superfluid localization transition in 1D.
In 
contrast to CDWs here the $\theta$-field may have vortex-like
singularities in space--time 
and the flow equations (\ref{eq:dK/dl.sg})-(\ref{eq:dw/dl.sg}) apply
again.  The vortex unbinding transition appears at $K_w$ with
$K_w^*=1/2, (q=2)$.  If both $w$ and $u$ are non-zero we can use the
canonical transformation to rewrite the vortex contribution in the
form $w\cos(q\varphi)$. For $K<K_u,K_w$ both perturbations are
irrelevant and the system is superfluid. For $K_w<K<K_u$ the decay of
$u$ is stopped due to the suppression of the $\varphi$ fluctuations
due to $w$. An Imry-Ma-argument shows further, that the
$q\varphi=\pi(n+1/2)$ state is destroyed on the scale $\xi\approx
\xi_w/u^2(\log\xi)$ by arbitrary weak disorder, i.e. vortices become
irrelevant above this scale. On larger scales one can expects that
quantum fluctuations wash out the disorder, the system is still superfluid.
Finally, at $K>K_u,K_w$ both perturbations are relevant and
superfluidity is destroyed.

To conclude we have shown, that in 1D CDWs/SDWs and superfluids
disorder driven zero temperature phase transitions are destroyed by
thermal fluctuations leaving behind a rich cross-over behavior.
Quantum phase slips in CDWs and superfluids lead to additional phase
transitions and shift the unpinning transition in CDWs to smaller
$K$-values.  Coulomb hardening and dissipative quantum effects will be
discussed in a forthcoming publication \cite{GlaNat}.

Acknowledgement: The authors thank A. Altland, S. Brasovskii,
T. Emig, L. Glazman, S. Korshunov, B. Rosenow and
S. Scheidl for useful discussions.

\end{document}